\documentclass[%
twocolumn,
superscriptaddress,
amsmath,amssymb,
aps,
prm,
floatfix,
]{revtex4-1}

\usepackage{amsmath}    
\usepackage{mathtools}
\usepackage{siunitx}
\usepackage{mhchem}
\usepackage[colorlinks=true,citecolor=blue]{hyperref}
\hypersetup{colorlinks=true,citecolor=blue,linkcolor=blue,urlcolor=blue}

\usepackage{tablefootnote}
\usepackage{dcolumn}  
\usepackage{bm}       
\usepackage[utf8]{inputenc}
\usepackage[T1]{fontenc}
\usepackage{mathptmx}
\usepackage{todonotes}
\usepackage{colortbl}
\usepackage[capitalise,noabbrev]{cleveref}

\usepackage{graphicx}   
\usepackage{verbatim}   
\usepackage{color}      
\usepackage{subfigure}  
\usepackage{hyperref}   
\newcommand{\etal}{\textit{et al.}}
\newcommand{\iasbs}{Department of Physics, Institute for Advanced Studies in Basic Sciences (IASBS), Zanjan 45137-66731, Iran}
\newcommand{\unibas}{Department of Physics, Universit\"{a}t Basel, Klingelbergstrasse 82, 4056 Basel, Switzerland}
\begin{document}

\title{Large Scale Structure Prediction of Near-Stoichiometric Magnesium Oxide Based on
a Machine-Learned Interatomic Potential: Novel Crystalline Phases and Oxygen-Vacancy Ordering}

\author{Hossein Tahmasbi}
\affiliation{\iasbs}
\author{Stefan Goedecker}
\affiliation{\unibas}
\author{S. Alireza Ghasemi}
\email{aghasemi@iasbs.ac.ir}
\affiliation{\iasbs}

\date{\today}
\begin{abstract}
    Using a fast and accurate neural network potential we are able to 
    systematically explore the energy landscape of large unit cells of 
    bulk magnesium oxide with the minima hopping method. The potential 
    is trained with a focus on the near-stoichiometric compositions, in 
    particular on suboxides, i.e., Mg$_x$O$_{1-x}$ with $0.50<x<0.60$. 
    Our extensive exploration demonstrates that for bulk stoichiometric 
    compounds, there are several new low-energy rocksalt-like structures 
    in which Mg atoms are octahedrally six--coordinated and form trigonal 
    prismatic motifs with different stacking sequences.
    Furthermore, we find a dense spectrum of novel non-stoichiometric crystal
    phases of Mg$_x$O$_{1-x}$ for each composition of $x$.
    These structures are mostly similar to the rock salt structure with 
    octahedral coordination and five--coordinated Mg atoms. 
    Due to the removal of one oxygen atom, the energy landscape becomes 
    more glass-like with oxygen-vacancy type structures that all lie 
    very close to each other energetically.
    For the same number of magnesium and oxygen atoms
    our oxygen-deficient structures are lower in energy if the vacancies 
    are aligned along lines or planes than rock salt structures with 
    randomly distributed oxygen vacancies. We also found the putative 
    global minima configurations for each
    composition of the non-stoichiometric suboxide structures. 
    These structures are predominantly composed of (111) slabs of the 
    rock salt structure which are terminated with Mg atoms at the top and 
    bottom, and are stacked in different sequences along the $z$-direction. 
    Like other Magn\'eli-type phases, these structures have properties 
    that differ considerably from their stoichiometric counterparts such 
    as low lattice thermal conductivity and high electrical conductivity.
\end{abstract}

\maketitle

\section{Introduction\label{sec:introduction}}
Magnesium oxide, one of the most abundant minerals in the lower mantle
of the earth~\cite{krauskopf1967introduction}, has been the subject 
of many experimental and theoretical studies due to its importance 
in various industrial applications~\cite{shand2006the}.
Furthermore, MgO has served as a prototypical system to evaluate the 
thermal conductivity of the deep mantle of the earth at high pressures
and temperatures~\cite{MacPherson_1982,Manga_1997,Beck_2007}, and  
can be used as a catalyst for various chemical reactions
~\cite{Trevethan_2007,Piskorz_2011,Pacchioni_2012}.
In particular, MgO can support metal clusters in catalytic processes
~\cite{Lu_2011,Aydin_2013}, and MgO clusters are good candidates for 
hydrogen adsorption due to the polarity of the Mg--O bonds
~\cite{Mojica_S_nchez_2019}. Also,
MgO surfaces with strong polarity, e.g. MgO(111), can be employed as 
support in photocatalytic water splitting experiments~\cite{Li_2019}.

Like some other binary solids such as sodium chloride and 
cadmium oxide, MgO crystallizes under ambient conditions in the cubic 
rock salt (RS) phase ($Fm\bar3m$), where the atoms are octahedrally 
coordinated. It has high thermal~\cite{Tang2010Mar} and low electrical 
conductivity~\cite{Sch_nberger_1995} with a large band gap of $7.83$~eV. 
Previous theoretical studies suggest that there exists a dense spectrum of 
low-energy polymorphs for MgO, some of which have been discovered and 
reported in the literature~\cite{Zwijnenburg_2010,Zwijnenburg2011}.
Therefore, the exploration of the potential energy landscape of MgO has 
attracted much attention in recent years
~\cite{Limpijumnong_2001,Sch_n_2004,Zwijnenburg2011,Stevanovi__2016}.
Limpijumnong~\etal~\cite{Limpijumnong_2001} studied
the transition between the wurtzite and 
RS phases of MgO and discovered the $h$-MgO polymorph with 
five-fold coordination.
The five-coordinated structures of MgO, together with the lowest-energy 
structures of ZnO and ZnS, have mainly been explored by 
Sch\"on~\cite{Sch_n_2004}. Zwijnenburg~\etal~\cite{Zwijnenburg2011} 
systematically mapped out the potential energy surfaces (PESs) of MgO 
using the basin hopping method in conjunction with
an interatomic potential implemented in the GULP code~\cite{Gale_2005}.
They discovered low--density phases of
MgO and found several new crystal structures with four-, five-, or 
six-coordinated Mg atoms which are relatively low in energy.
Furthermore, the PES of MgO was explored by a novel 
approach which uses the random superlattice structure sampling
~\cite{Stevanovi__2016} followed by local relaxations at the density 
functional theory (DFT) level. It resulted in the discovery 
of different polymorphs of several ionic systems, e.g. MgO, ZnO, and SnO$_2$, 
and the experimental structures were found more frequently than 
the hypothetical structures.
In fact, the surprising complexity of the PESs of binary solids that 
crystallize in octahedral structures at ambient conditions makes it 
difficult to explore in an exhaustive way the spectrum of 
possible structures.

Only little is known about the PESs of non-stoichiometric crystal 
phases of MgO (i.e. Mg$_n$O$_m$, $n\neq m$) since most of the 
previous studies have focused on the stoichiometric composition.
However, many efforts have been made to better understand the defects 
at the bulk and the surface of MgO~\cite{Ferrari_1995}. Among all the defects,
the oxygen vacancy is the most important that can substantially change
properties and the chemical behaviour of MgO 
structures~\cite{Pacchioni_2003,Pacchioni_2012}. 
In addition, a few attempts have been made to study non-stoichiometric
clusters of MgO~\cite{Roberts_2001,Uchino_2012}. 
Such studies have demonstrated that non-stoichiometric clusters exhibit 
peculiar magnetic properties~\cite{Bhattacharya_2013}.

In recent years, neural network potentials (NNPs)~\cite{Behler2007,Ghasemi2015} 
have emerged to overcome computational limitations of density functional 
based approaches. Because of its reduced numerical cost, this approach 
makes it possible to model the PESs of large systems with hundreds or even 
thousands of atoms. In this way new phenomena that can only be observed on 
larger length scales can be discovered. Vacancy ordering is such a phenomenon 
that will be investigated in this study.
Neural network potentials have already been successfully constructed 
for different compounds and have been employed for crystal structure predictions
~\cite{Faraji2017,Rasoulkhani2017,Hafizi_2017,Rostami2018,Eshet_2010}, 
to discover 2D materials~\cite{Eivari2017}, 
to study gas-surface interactions~\cite{Shakouri_2017}, 
for the systematic investigations of surface reconstructions~\cite{Faraji_2019}, etc. 

In this work, we train a NNP to study the MgO system based on 
the Charge Equilibration via Neural network Technique 
(CENT)~\cite{Ghasemi2015} as implemented in the FLAME code~\cite{Amsler2020}.
A more detailed description of CENT can be found elsewhere
~\cite{Ghasemi2015,Faraji2017,Amsler2020}.
We demonstrate for the first time that the CENT 
method can also accurately reproduce the PES of non-stoichiometric 
systems that have a more conmplex potential energy surface. 
For this purpose, we train a highly transferable neural network 
potential for magnesium oxide for both stoichiometric 
and non-stoichiometric compositions of clusters and crystals. 
We show that this potential gives rise to accurate second and third-order 
interatomic force constants which allow us to compute dynamical properties 
that are in good agreement with DFT results.

We use this NNP to probe the energy landscape of bulk structures 
of MgO with a main focus on the suboxide near-stoichiometric compositions
which are of fundamental importance.
For the stoichiometric compounds, 
we find novel low-energy polymorphs for MgO in addition to 
recovering all the well-known structures in the literature.
To represent these structures, 
large simulation cells containing up to $32$ atoms are required.
For the near-stoichiometric compositions of MgO, namely, 
Mg$_x$O$_{1-x}$ with $0.51\leq x\leq0.57$,
we find many new polymorphs with similar structural features
which mostly have small band gaps or pseudogaps due to the appearance 
of defect states in the gap. They are energetically more favorable than
the classical defect structures of RS. 

In addition, we find that the global minimum of all compositions have 
nearly identical structural motifs and symmetries.
These structures are RS structures along the $[111]$ direction
in which a layer of oxygen atoms is removed. Therefore,
they are composed of slabs which terminate with Mg atoms and
only the number of their layers or the octahedral motifs changes
with the number of atoms in their primitive cell.
Unlike RS, these structures have metallic behaviour with low
lattice thermal conductivity.

This manuscript is organized as follows. In Sec.~\ref{sec:method} 
we describe the methods employed in this work including the training 
process of the CENT potential and its validation as well as 
our search method. Sec.~\ref{sec:result} contains the results on 
the stoichiometric and non-stoichiometric bulk structures of MgO. 
Finally, the main conclusions are summarized in Sec.~\ref{sec:conclusion}.

\section{Method} \label{sec:method}

\subsection{Density functional theory}
We employ two different software packages to perform the DFT 
calculations in this work.
The training data for our CENT potential are generated using  
the \textit{ab initio} molecular simulation package (FHI-aims)~\cite{Blum_2009}. 
We use the PBE exchange-correlation functional~\cite{Perdew_1996pbe}
for both free and periodic boundary conditions with the default tight settings, 
i.e. tier $1$ and tier $2$ basis functions for Mg and O, respectively.
The predicted structures from the NNP and their properties (e.g., geometry relaxation,
phonon dispersions etc.) are refined using the projector augmented
wave (PAW) formalism as implemented in the Vienna \textit{ab initio} 
Simulation Package (VASP)~\cite{Kresse1993,Kresse_1994,Kresse_1996,Kresse_1996_prb}.
A plane-wave cutoff energy of $550$ eV and k-point mesh with a density 
of $0.03$~$\AA^{-1}$ were used to obtain converged results.

\subsection{Neural network potential}

\subsubsection{Training} \label{subsec:train}
To probe the PESs of the near-stoichiometric compounds, 
we train an accurate artificial neural network (ANN) potential 
to model atomic interactions using the CENT method.
Our ANN potential models the PESs of both stoichiometric and various 
non-stoichiometric MgO systems with the same parameter set. 
This is highly challenging since we have to accurately describe 
not only the configurational space but also the compositional space.
For this purpose, the CENT potential has to be trained with a hierarchical 
approach, and a variety of different reference data sets are required.

To start the training process, we use NaCl-type structures which were 
obtained from Ref.~\cite{Ghasemi2015} as the initial training 
data set by scaling the bond lengths to typical values for MgO.
Also, we include the global minima (GM) structures of MgO clusters of 
different sizes (MgO)$_{4}$ to (MgO)$_{36}$ which were found 
by Chen~\etal~\cite{Chen2014}. 
As a result, we select more than $2200$ diverse structures to 
train a CENT potential in the first step.
The ANN architecture $70-5-5-1$ is employed, i.e., $70$ symmetry functions, 
including $16$ radial functions and $54$ angular functions~\cite{Behler2011}, 
with two hidden layers each containing $5$ nodes, and one output layer.
The output layer is taken as the electronegativity of the given atom 
where the input layer is fed with the symmetry functions as
environment descriptors with a cutoff radius of $11$~Bohr.

By combining this first, initially trained potential together with the 
minima hopping (MH) method (see Sec.~\ref{sec:mh}), we perform a 
preliminary search on the high dimensional PES of (MgO)$_n$ clusters
with $n = 8, 11, 13, 14, 19, 26, 28,~\text{and}~40$.
After removing all similar and high energy structures by 
comparing the environment descriptors~\cite{Behler2011,Ghasemi2015} 
implemented in FLAME, we select more than $7100$ stoichiometric neutral clusters 
with a large diversity. This filtering step is crucial to prevent overfitting.
Using this data set, we then train in a second step a CENT potential 
that can be applied to the stoichiometric bulk and cluster structures of MgO.
Note that twenty percent of the data set is randomly selected 
as a validation set to monitor the performance of the training.

Using this second CENT potential, we explore the PES
of stoichiometric clusters (MgO)$_n$ ($n=4-40$) and crystalline
Mg$_n$O$_n$ ($n=2-48$) structures.
We recover all GM structures which were previously reported
for clusters of magnesium oxide
~\cite{Aguado_2000,Roberts_2001,Haertelt_2012,Chen2014}.
Further, we use this second potential to find several non-stoichiometric 
MgO structures, i.e., Mg$_x$O$_{1-x}$ for both clusters 
($x = 0.53$ to $x = 0.60$) and bulk systems ($x=0.55$ to $x=0.57$) 
with different supercell sizes. 
In this way, we find a diverse set of non-stoichiometric structures although
non-stoichiometric structures had not been included in the training data set.

In a third, final step, we expand the reference data set by adding
all structures, both stoichiometric and non-stoichiometric,
that we generate in step two to the training set. 
In total, we use $24180$ structures as training data set 
including both stoichiometric and non-stoichiometric clusters and 
crystalline structures. 

Note that in each step of the training process, 
we run the FLAME code several times with different initializations  
of both the CENT parameters
(electronegativity offset, hardness, Gaussian widths) and neural network weights.
The best parameter set for the final calculations is selected based on the root 
mean squared error (RMSE) values of energy and atomic forces as well as
the size of the variations in charge and electronegativity.
In the final training step, the RMSEs of energy and atomic forces are less than 
$11$~meV/atom and $0.257$~eV/\AA, respectively.
The parameters and weights of the trained potential for MgO are available
in the FLAME code~\cite{Amsler2020}.

\subsubsection{Validation}\label{subsec:valid}
\begin{figure}[ht]
    \centering
        \includegraphics[width=1.00\columnwidth]{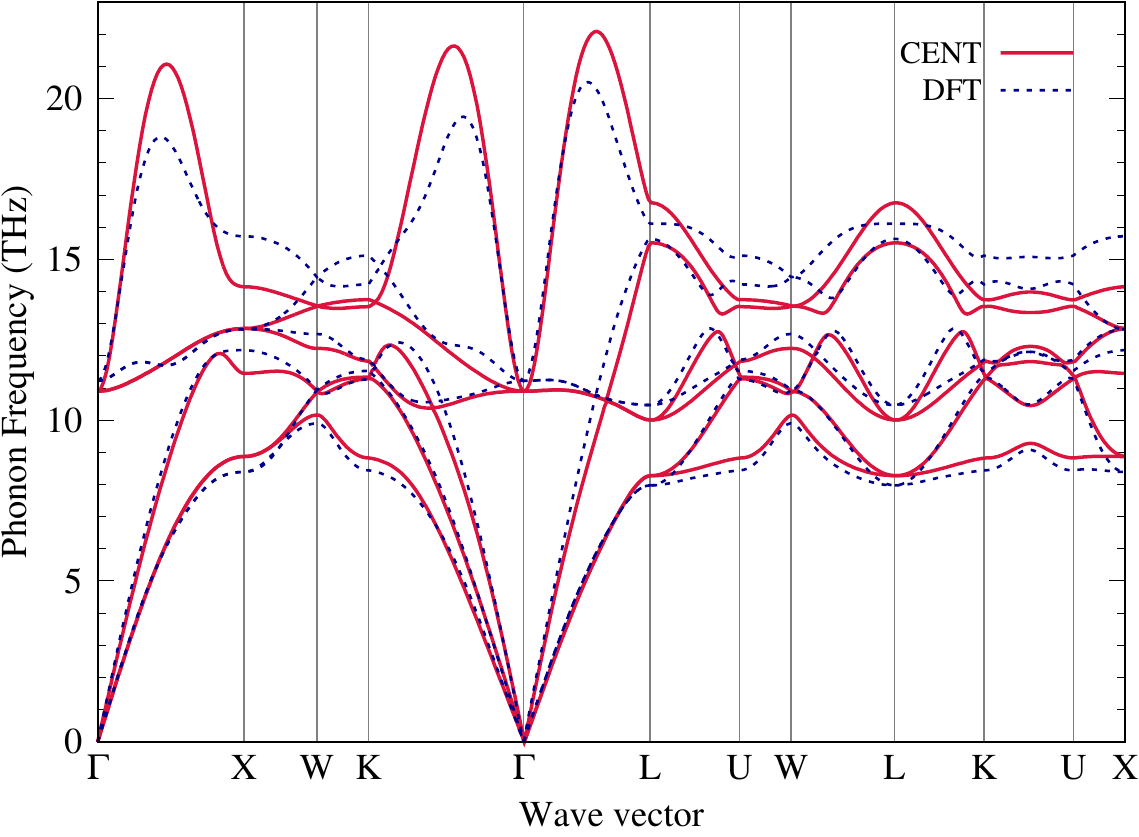}
    \caption{Comparison of the phonon dispersion of the MgO rock salt obtained 
    with DFT and CENT calculations. The LO-TO splitting was neglected.}
    \label{fig:Ph_vasp_cent}
\end{figure}

In order to examine the quality and reliability of the CENT potential, 
we employ it to calculate various (physical) properties. 
As a first test,
we perform phonon calculations for the RS structure with the frozen phonon 
approach as implemented in the PHONOPY package~\cite{Togo2015}.
A large supercell $5\times5\times5$ containing 250 atoms is used to ensure 
the convergence of the force constants and the phonon density of states (DOSs). 
\cref{fig:Ph_vasp_cent} compares the phonon dispersions calculated by DFT 
and CENT, showing good agreement in particular for the acoustic modes, 
which dominate the thermal transport properties.
Since the optical branches have very small group velocities, 
they contribute little to the heat transport process~\cite{Morelli,Slack_1973}. 

Next, we compute the lattice thermal conductivity of 
the RS phase of MgO, which requires also 
the third-order derivatives of the potential energy.
We compute the second and third-order interatomic force constants (IFCs)
with PHONOPY and thirdorder.py~\cite{Li2014Jun}, respectively. 
The third-order IFCs are calculated with $5\times5\times5$ supercells,
truncating the interactions beyond the sixth nearest neighbors.
The IFCs are fed into the ShengBTE code~\cite{Li2014Jun} to 
iteratively solve the Boltzmann transport equation for phonons.

\begin{figure}[ht]
    \centering
        \includegraphics[width=1.00\columnwidth]{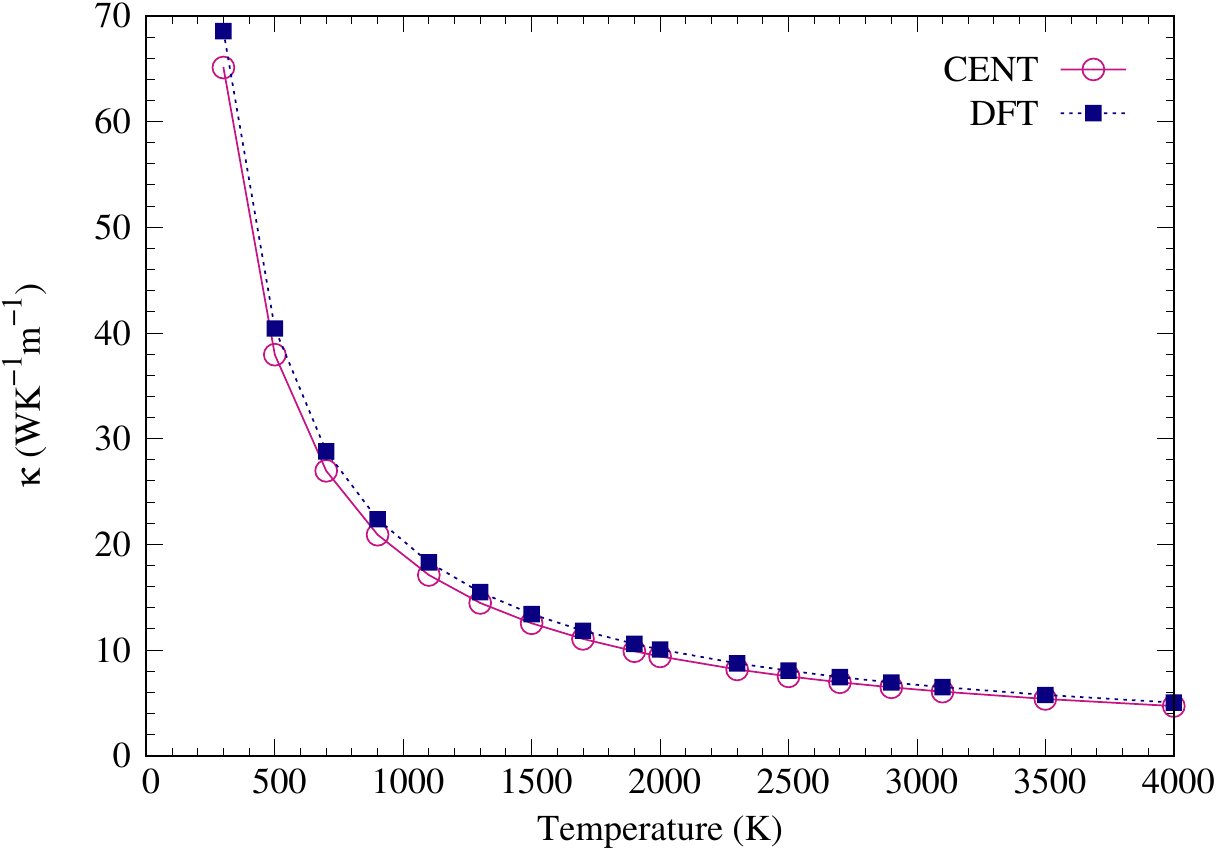}
    \caption{Comparison of the DFT and CENT results for the thermal conductivity of 
    the MgO rock salt at different temperatures.}
    \label{fig:TC_vasp_cent}
\end{figure}

\cref{fig:TC_vasp_cent} shows the lattice thermal conductivity
of RS at different temperatures using DFT and CENT.
The comparison shows good
agreement between the DFT and CENT, with deviations
between $5$\% and $7$\% at $300$~K and $4000$~K, respectively. 
Further, our results are in excellent agreement with the experimental 
and theoretical results in the literature
~\cite{Tang2010Mar,Lindsay2015Mar,Dekura_2017}.

Finally, 
we use the MH method to screen the PESs of both cluster and crystal 
structures with different sizes and stoichiometries.
We validate that our CENT potential reproduces all results in 
the literature~\cite{Chen2014,Schleife2006,Zwijnenburg2011}
as local minima on the PESs.
Furthermore, we compare the energetics of the local minima from CENT with 
reference DFT values to ensure that the energetic ordering agree with 
each other in almost all cases. 
We also repeated the phonon calculations for several of the non-stoichiometric structures,  
and performed geometry relaxation for $100$ stoichiometric and non-stoichiometric bulk 
structures to compare their energies with DFT values.
Based on these rigorous tests, we conclude that our ANN potential is sufficiently 
accurate to perform structure prediction as well as to accurately estimate 
the thermal properties of the MgO system.

\subsection{Structural search\label{sec:mh}}
We explore the energy landscape of bulk phases of MgO 
in a systematic way using the minima hopping (MH) method
~\cite{Goedecker2004,Amsler2010,MHM3,MHM4}
which is an efficient search approach for global optimization and 
is implemented in the FLAME code.
We investigate stable phases of Mg$_x$O$_{1-x}$ 
with $x=0.51$ to $0.57$, i.e., for the seven compositions Mg$_4$O$_3$ ($x=0.57$), 
Mg$_5$O$_4$ ($x=0.56$), Mg$_6$O$_5$ ($x=0.55$), 
Mg$_7$O$_6$ ($x=0.54$), Mg$_8$O$_7$ ($x=0.53$), 
Mg$_{12}$O$_{11}$ ($x=0.52$), and Mg$_{20}$O$_{19}$ ($x=0.51$). 
Our simulation cells contain up to 22 f.u., i.e., supercells 
including $7$ to $46$ atoms.

Even though the stoichiometric bulk
structures Mg$_n$O$_n$ ($n = 4$ to $16$, $x=0.50$) were 
originally intended only as a validation test, 
we discovered a plethora of new,
low energy structures for these stoichiometric configurations when using
larger unit cells. For each simulation cell size, 
we perform MH runs at least $20$ times with different 
starting configurations to thoroughly scan the PES.

Selected structures are then refined at the DFT level. 
In this selection process, we first remove all structures with CENT energies higher 
than $300$~meV/atom with respect to the minimum structure in each composition.
Then, we remove duplicate structures by comparing them based on 
their energies and space groups, i.e., two structures with the same space group are 
considered distinct if the difference in their CENT energies is 
larger than $10^{-4}$~Ha.
\section{Results and discussion\label{sec:result}}

\subsection{Stoichiometric structures}\label{sec:stoich}
We carry out structure prediction simulations for the stoichiometric 
phases of MgO with the minima hopping method.
During this search we discover metastable phases
and some of their stacking faults which have been previously reported 
either in materials databases~\cite{Jain2013,Saal_2013} 
or in the literature~\cite{Chen2014,Schleife2006,Zwijnenburg2011}. 
The structural data of these polymorphs are presented in 
the Supplementary Material~\cite{sup_mat}.
We enumerate these structures with labels from S01 to S25, 
and for each one of them we list the corresponding space-group, 
the number of atoms in their unit cell, 
the relative formation energies ($\Delta$E$_f$), 
the volumes per Mg atom (V/V$_{RS}$) with respect to
the ground-state RS structure, and the band-gap energies 
in \cref{tab:st_mgo}.
We define the formation energy per atom, 
\begin{equation}
E_f = (E(Mg_nO_m)-nE(Mg)-\frac{m}{2}E(O_2))/(n+m),
\label{eq:Ef}
\end{equation}
where E(Mg$_n$O$_m$) is the total energy of the structure,
and $n$ and $m$ are the number of magnesium and oxygen atoms, respectively.
Also, E(Mg) is the energy of a single magnesium atom, and E(O$_2$) 
is the energy of an isolated oxygen molecule in its triplet ground state.

\begin{table*}
\centering
\begin{ruledtabular}
    \caption{Structural data of stoichiometric phases of MgO. 
Column 1--3 contain the label, the phase name, and the space group, respectively.
Column 4 contains the number of atoms in the unit cell of Mg$_n$O$_n$. 
Columns 5 and 6 contain the relative formation energy per atom with respect to 
    the RS phase and the volume per Mg atom, respectively. 
    The last column contains the PBE band gap.}
\label{tab:st_mgo}
\begin{tabular}{lclcccl}
Label & Phase  &  Space group  & N &  $\Delta$E$_f$ (eV/atom) & V/V$_{RS}$  & Gap (eV)\\
\colrule
 S01  &RS                 & $Fm\bar{3}m$ (225)    & 8 &  0.000   & 1.000 &4.45 \\
 S02  &                   & $P\bar{3}m1$ (164)    &24 &  0.026   & 1.003 &4.13 \\
 S03  &                   & $R3m$ (160)           &24 &  0.033   & 1.004 &3.75 \\
 S04  &$h$-MgO              & $P6_3/mmc$ (194)      & 8 &  0.038 & 1.183 &3.26 \\
 S05  &                   & $P6_3/mmc$ (194)      &24 &  0.040   & 1.006 &4.03 \\
 S06  &                   & $R\bar{3}m$ (166)     &24 &  0.051   & 1.006 &4.15 \\
 S07  &                   & $R3m$ (160)           &24 &  0.065   & 1.009 &3.72 \\
 S08  &                   & $P3m1$ (156)          &12 &  0.065   & 1.009 &3.72 \\ 
 S09  &Wurtzite           & $P6_3mc$ (186)        & 8 &  0.074   & 1.267 &3.38 \\
 S10  &                   & $R\bar{3}m$ (166)     &32 &  0.076   & 1.009 &4.12 \\    
 S11  &                   & $Ibam$ (72)           &32 &  0.076   & 1.233 &3.07 \\    
 S12  &                   & $R3m$ (160)           &12 &  0.087   & 1.273 &3.43 \\    
 S13  &                   & $Cmm2$ (35)           &14 &  0.087   & 1.153 &3.27 \\    
 S14  &                   & $Pbcm$ (57)           &24 &  0.088   & 1.256 &3.02 \\
 S15  &Stacking fault of Wurtzite & $P6_3mc$ (186)& 8 & 0.092    & 1.274 &3.44 \\
 S16  &                   & $P4_2/mnm$ (136)      & 8 &  0.094   & 1.326 &3.36 \\
 S17  &                   & $R\bar{3}m$ (166)     &12 &  0.100   & 1.011 &4.13 \\
 S18  &                   & $I4/mcm$ (140)        & 8 &  0.101   & 1.328 &2.92 \\
 S19  &Zincblende         & $F\bar{4}3m$ (216)    & 8 &  0.107   & 1.276 &3.46 \\
 S20  &                   & $P6_3/mmc$ (194)      & 8 &  0.149   & 1.015 &4.12 \\
 S21  &                   & $Pnma$ (62)           & 8 &  0.152   & 1.434 &3.17 \\
 S22  &                   & $Cmc2_1$ (36)         & 8 &  0.156   & 1.093 &3.64 \\
 S23  &                   & $I4/m$ (87)           & 8 &  0.160   & 1.509 &3.29 \\
 S24  &                   & $I\bar{4}3m$ (217)    & 8 &  0.204   & 1.597 &3.04 \\
 S25  &                   & $Fddd$ (70)           & 8 &  0.222   & 1.682 &3.01 \\
\end{tabular}
\end{ruledtabular}
\end{table*} 

In addition to the well-known phases like the RS, $h$-MgO, wurtzite, and zincblende,
which are the ambient ground states of many binary compounds such as ZnO and ZnS
~\cite{Sangthong_2010,Rasoulkhani2017,Zwijnenburg2011}, 
our MH runs reveal several new low energy polymorphs 
with 12 and 24 atoms per unit cells (i.e. Mg$_6$O$_6$ and Mg$_{12}$O$_{12}$).
Like in the RS structure, the Mg atoms in these phases have only octahedral bonding.
In contrast, the energetically higher wurtzite and zincblende structures are 
composed of MgO$_4$ tetrahedrons which are stacked in hexagonal sequences.
Therefore, unlike in some binary oxides like ZnO, 
octahedral bonding is the energetically preferred structural motif for MgO. 

In particular, we discover two new phases of MgO, namely S02 and S03,
with 24 atoms per unit cell in the energy range of $38$ meV/atom, between RS and $h$-MgO.
Earlier systematic structural searches for MgO bulk structures
~\cite{Stevanovi__2016,Zwijnenburg2011} with unit cells up to 40 atoms 
seem to have missed these two phases, and,
to the best of our knowledge, we are the first to report them here.
Their small energy differences with respect to the RS ground state
($26$ and $33$~meV/atom, respectively) suggests that these polymorphs are well within the 
synthesizability limit of oxide materials~\cite{Sun_2016}.

\begin{figure*}
    \centering
        \includegraphics[width=2.00\columnwidth]{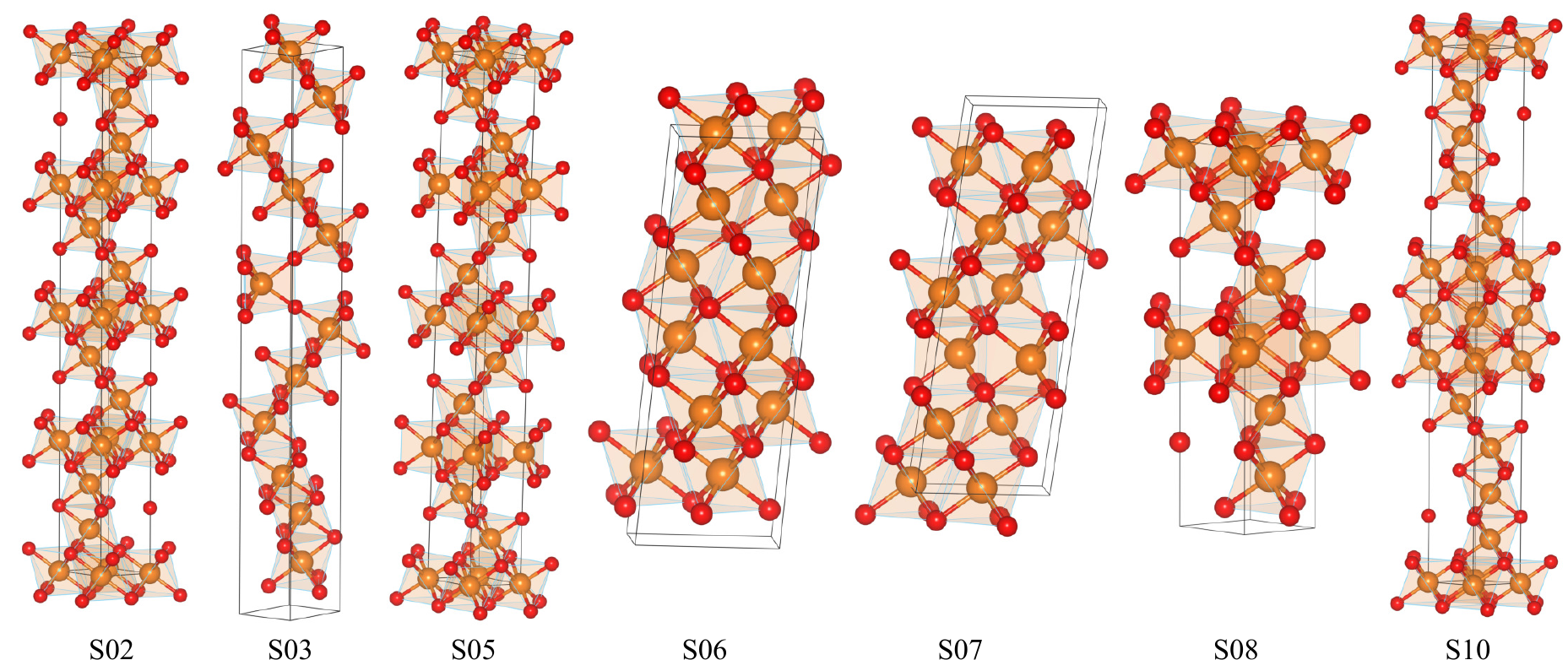}
    \caption{The 7 low-energy modifications of the RS phase.
    Orange (large) and red (small) spheres denote Mg and O atoms, 
    respectively.
    The structures S05~\cite{Zwijnenburg2011} and 
    S10~\cite{Stevanovi__2016} are reported 
    in the literature and the others are new based on this work.}
    \label{fig:bulk_st}
\end{figure*}

From the structural perspective, these two new polymorphs are modifications 
of the RS structure and are composed of MgO$_6$ 
octahedra and trigonal prisms, similar to the NiAs structure type, 
which are stacked on top of each other in various sequences and directions.
Other structures with similar structural features, like the S10 and S05 
(another well-known rocksalt-like structure~\cite{Zwijnenburg2011, Saal_2013}), 
have higher energies.
In \cref{fig:bulk_st}, we show the structures of
the $7$ low-energy MgO phases of MgO which are modifications of rock salt.
As shown in \cref{fig:bulk_st}, S02 like RS is composed of the same octahedra 
with the difference that their directions or connectivities 
are changed twice in its unit cell.
In fact, the connectivity of the octahedra is changed from edge-sharing
to face-sharing where they are flipped. 

\begin{figure}[ht]
    \centering
        \includegraphics[width=1.00\columnwidth]{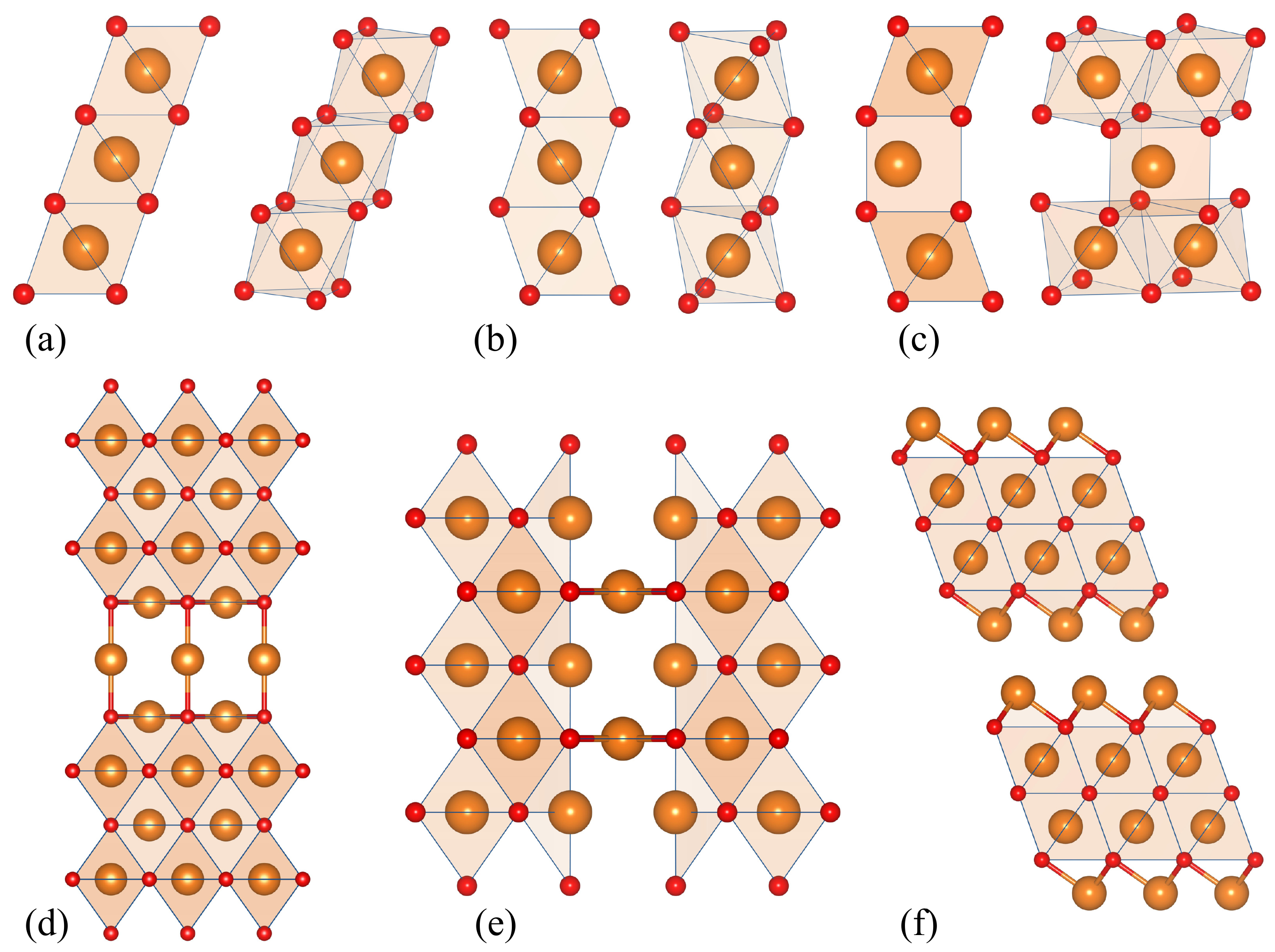}
    \caption{(a)-(c) Motifs of the near-stoichiometric MgO structures.
    (d) NS3-03, (e) NS5-10, and (f) NS7-01 are 
    three examples of the non-stoichiometric phases:
    (d) and (e) show the similarity in the non-stoichiometric structures 
    in which lines of oxygen atoms are removed. (f) shows one of the GM 
    structures in which a layer of oxygen atoms is removed.}
    \label{fig:polyhedra}
\end{figure}

The different structures that we observe in \cref{fig:bulk_st} 
arise from subtle changes in the arrangements of the \ce{MgO6} octahedra.
The possible arrangements that we observe are shown schematically in 
\cref{fig:polyhedra}(a)--(c), in 2D and 3D representations. 
The octahedra in the RS structure are all edge-connected, 
as shown \cref{fig:polyhedra}(a). \cref{fig:polyhedra}(b)
indicates that when an octahedron is flipped in the second layer 
then two consecutive octahedron will share a triangular face of the octahedra.
In the S02 unit cell, it can be seen that this flip is happened
twice. Therefore, S02 is composed of two layers of the RS with 
the stacking sequence \textit{AB}.
For S03, in addition to the octahedra flip, 
we also see trigonal prisms which have edge-sharing with the octahedra 
as \cref{fig:polyhedra}(c) represents.
Therefore, the stacking sequence for S03 becomes \textit{ABC}.
Besides these structures, we discovered three new low 
energy rocksalt-like polymorphs i.e. S06, S07, and S08 which are 
energetically more favorable than the well-known structure
S11 with symmetry $Ibam$. \cref{fig:bulk_st} shows that S06 is 
similar to S02 with the identical stacking 
sequence \textit{AB} of octahedra.  Structures S07, S08 and S03 share the 
stacking sequence \textit{ABC}, but are composed 
of octahedra and trigonal prisms in different orderings.
Also, note that the stacking periodicity for the well-known 
structure S05 is \textit{ABCD}.
Along the $z$-direction, these rocksalt-like structures are identical to 
RS in the $[111]$ direction.

\begin{figure}[ht]
    \centering
        \includegraphics[width=1.00\columnwidth]{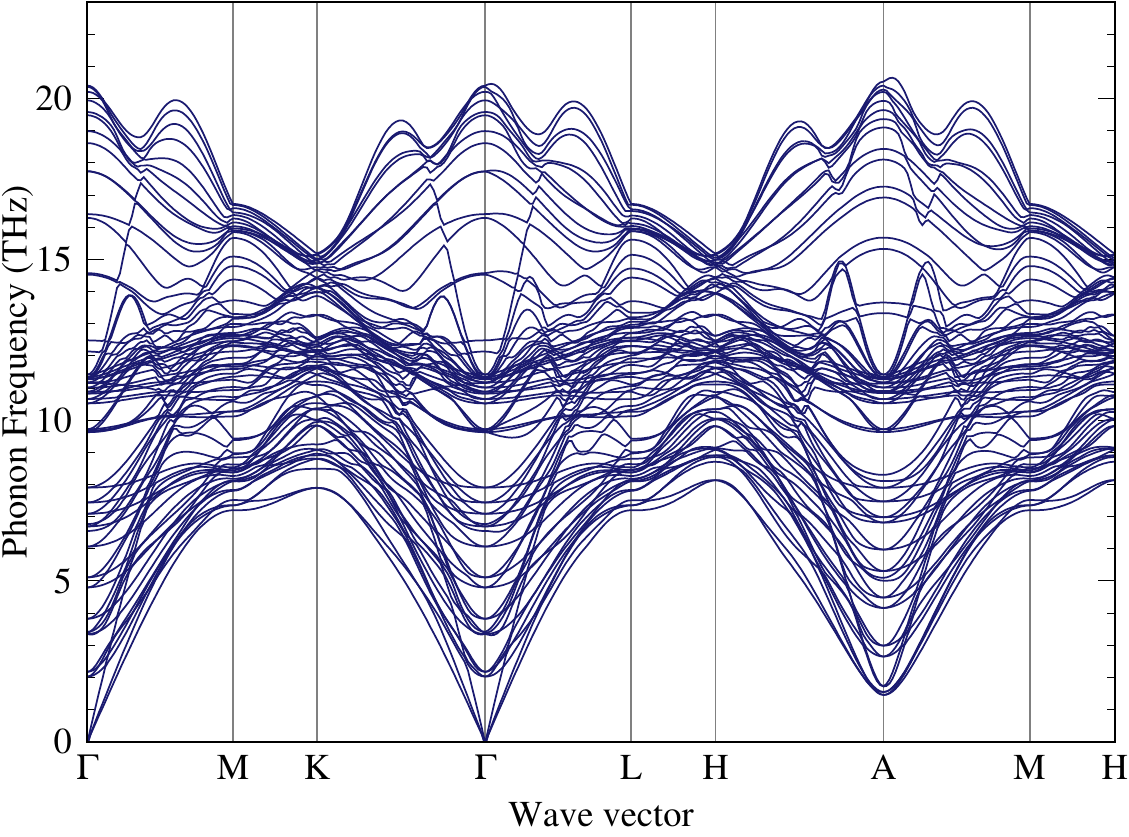}
    \caption{The PBE phonon dispersion of S02, a new stoichiometric phase of MgO.
    The LO-TO splitting is neglected.}
    \label{fig:ph_S02}
\end{figure}

The phonon dispersions for all the stoichiometric structures are calculated
to evaluate the dynamical stability. Supercells of dimensions 
$2\times2\times2$ up to 
$5\times5\times5$ were used depending on the unit cells.
For instance, \cref{fig:ph_S02} shows the phonon dispersion of S02.
Our phonon calculations reveal that $h$-MgO phase (S04) 
is dynamically unstable at ambient conditions in agreement 
with earlier studies~\cite{Sch_n_2004,Zwijnenburg2011}.

The electronic band structures of the stoichiometric phases of MgO  
were also calculated at the level of the PBE functional. 
As the results in \cref{tab:st_mgo} indicate,
all these phases have wide band gaps.
The new polymorphs S02, S03, S06, S07, and S08 which are 
modifications of RS with almost the same densities, 
have also a wide direct band gap which is however a little bit smaller than 
in the RS phase. 

\subsection{Structures and properties of non-stoichiometric polymorphs\label{sec:nonst}}

\begin{figure}[ht]
    \centering
        \includegraphics[width=1.00\columnwidth]{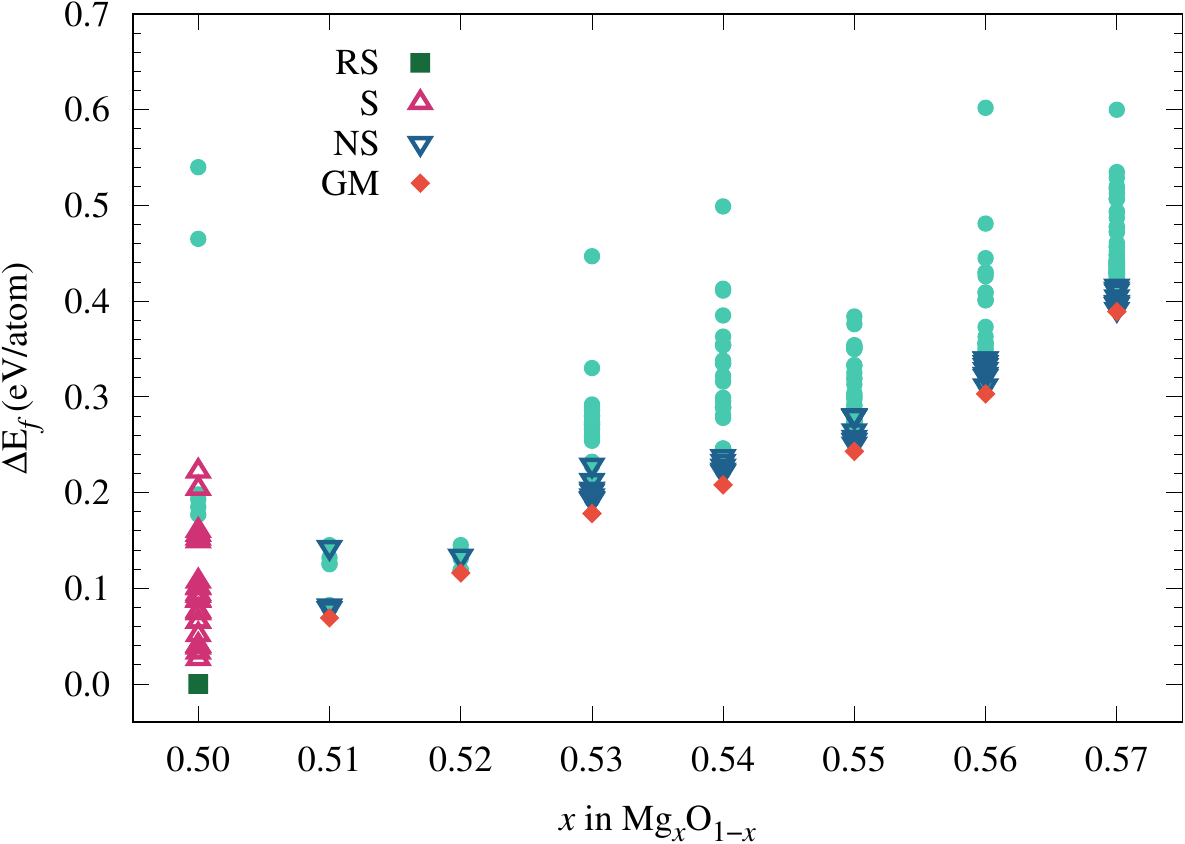}
    \caption{The phase diagram of Mg-O near the stoichiometric 
    region. The formation energy differences $\Delta$E$_f$, calculated from 
    \cref{eq:Ef}, show the distances 
    from the convex hull as a function of $x$. 
    The points S and NS represent the structures
    in \cref{tab:st_mgo} and \cref{tab:nonst}, respectively. 
    The GM points show the global minimum structure of each composition $x$. 
    The green filled circles also show all 
    the other structures which are found in this work.}
    \label{fig:energy_x}
\end{figure}

In the so-called near-stoichiometric region of Mg$_x$O$_{1-x}$
where $x$ varies between $0.51$ and $0.57$, 
$471$ structures where retained and validated using DFT. 
We find many new polymorphs 
with formation energies that are up to $\approx$~600~meV/atom 
higher than the RS phase.
\cref{fig:energy_x} shows the energetics of these novel polymorphs 
compared to the stoichiometric phases (\cref{tab:st_mgo}).

These structures are similar to each other with regard to their energies 
and space groups for a given composition. 
Hence, out of the many polymorphs that we discover for each composition, 
only $48$ structures are selected to study their dynamical and electronic
band properties in detail.
(see Supplementary Material~\cite{sup_mat} for their structural data).
The key properties of these non-stoichiometric structures are given 
in \cref{tab:nonst}.

We also compare the energy of these structures with their corresponding 
rock salt structures with defects for each composition of $x$.
To this end, we first calculate the energy of an oxygen vacancy in the 
RS structure:
\begin{equation}
    E_v = E_{RS}(Mg_kO_{k-1}) - E_{RS}(Mg_kO_k) + \frac{1}{2}E(O_2),
\end{equation}
where E$_{RS}$(Mg$_k$O$_k$) is the energy of a cubic rock salt structure 
with $k=108$ and E$_{RS}$(Mg$_k$O$_{k-1}$) is the energy of the cubic rock salt 
with one oxygen vacancy.
We can then define a reference energy for each composition in which 
the interaction of defects is ignored,
\begin{equation}
    E_{ref}(Mg_nO_m) = E_{RS}(Mg_nO_n) + (n-m)E_v - \frac{(n-m)}{2}E(O_2), 
\end{equation}
where $E_{ref}(Mg_nO_m)$ is the energy of a RS structure with random 
oxygen defect(s) and without considering the interaction between defects.
Therefore the relative defect energy per atom $\Delta$E$_d$ is given by,
\begin{equation}
    \Delta E_d = (E(Mg_nO_m) - E_{ref}(Mg_nO_m))/(n+m),
\end{equation}
where $(E(Mg_nO_m)$ is the total energy of the non-stoichiometric structures
and the number of removed oxygen atoms $(n-m)$ varies between $1$ to $4$ 
in the structures. Although for some non-stoichiometric structures with 
a small unit cell $E_{ref}(Mg_nO_m)$ could be less accurate without 
the interaction effects of defects, our results (see~\cref{tab:nonst}) 
indicate that most of the structures are more favorable than 
a RS structure with defects.

Note that for each composition $x$, we find similar configurations 
with different symmetries whose energies lie in a very narrow band. 
This indicates near-degeneracy in the non-stoichiometric structures. 
For example, the structure NS3-05 was selected from eight configurations
with similar structural features that had however different space groups. 
As another example of a nearly degenerate case, we have selected 
the structures NS5-07--NS5-09 that have nearly identical formation 
energies and almost the same symmetries but different structures.
Except for the structure NS3-08 which is similar to $h$-MgO, 
all the other non-stoichiometric structures are indeed similar 
to the RS structure. 
However, the values of the ratio of volume per 
Mg atom with respect to RS indicate that the cell does not shrink 
by the removal of one or several oxygen atoms but that, 
on contrary, it expands slightly. 

These structures are formed by removing a layer of oxygen atoms located
on a $(111)$ plane or a line of oxygen atoms in the $<110>$ direction in 
the RS structure as depicted schematically in \cref{fig:polyhedra}(d-f) 
for three of them. In other words, we see that the oxygen vacancies are 
ordered along lines or in planes. 

To better understand the nature of these structures, 
we calculated the energy of the cubic RS structures 
Mg$_{216}$O$_{216-l}$ ($l=2,3$), with two or 
three oxygen vacancies in a perfect cell of $512$ atoms.
Considering all possible arrangements of the two vacancies, we found 
that the nearest-neighbor double vacancy along the $<110>$ is the lowest 
in energy by $~80$ meV.
Adding a third vacancy at various places gives the lowest energy 
if they are adjacent in the $(111)$ plane. This clustering lowers 
the energy by $~235$ meV compared to configurations with large distances 
among the vacancies.
These fairly strong interactions between vacancies explain the regular 
large scale patterns that we found for the non-stoichiometric structures. 
In fact, these results confirm the existence of oxygen-vacancy ordering, 
which has experimentally been observed 
~\cite{Scott_2000,Torbr_gge_2007,Murgida_2014}
but to the best of our knowledge 
never been theoretically found by a structure prediction. 
It is worth mentioning that vacancy ordering can also affect the physical 
properties of materials such as electrical and thermal 
properties~\cite{Zhang_2012,Siegert_2014}.
Although we use random structures with different concentrations of oxygen 
vacancies as initial structures, our MH runs with the CENT potential can
predict the oxygen vacancy ordering, extending in this way the power of 
systematic structure prediction to length scales larger than the size of 
a crystalline unit cell.

\begin{table*}
\centering
    \caption{Structural data of non-stoichiometric phases of MgO. 
Column 1 contains label, followed by the space group.
Columns 3--4 contain $x$ values and number of atoms in 
    the unit cell of Mg$_n$O$_{m}$, respectively. 
Column 5 and 6 contain the relative defect and formation energies per atom, respectively.
Column 7 lists the ratio of volume per Mg atom. 
The last column contains the PBE band gap (* the structures with a pseudogap).}
\label{tab:nonst}
\begin{ruledtabular}
\begin{tabular}{clclrccc}
Label&Space group& $x$  & N (n, m) & $\Delta$E$_d$ (eV/atom)& $\Delta$E$_f$ (eV/atom)&V/V$_{RS}$&Gap (eV) \\
\colrule
NS1-01 &  $P\bar{3}m1$ (164)   & 0.51 &39 (20, 19)   &  $-0.009$& 0.069	& 1.005 & *    \\ %
NS1-02 &  $Cmmm$ (65)          & 0.51 &39 (20, 19)   &  $ 0.001$& 0.079	& 1.005 & 1.18 \\ %
NS1-03 &  $Immm$ (71)          & 0.51 &39 (20, 19)   &  $ 0.004$& 0.082	& 1.006 & 2.35 \\ %
NS1-04 &  $R3m$ (160)          & 0.51 &39 (20, 19)   &  $ 0.065$& 0.143	& 1.015 & 1.77 \\ %
\arrayrulecolor{gray}\hline
NS2-01 &  $R\bar{3}m$ (166)    & 0.52 &46 (24, 22)   &  $-0.016$& 0.116	& 1.009 & *    \\ %
NS2-02 &  $Cmmm$ (65)          & 0.52 &23 (12, 11)   &  $ 0.001$& 0.134	& 1.009 & 0.82 \\ %
\hline
NS3-01 &  $P\bar{3}m1$ (164)   & 0.53 &30 (16, 14)   &  $-0.025$& 0.178	& 1.014 & *    \\ %
NS3-02 &  $C2/m$ (12)          & 0.53 &30 (16, 14)   &  $-0.011$& 0.193	& 1.013 & 0.81 \\ %
NS3-03 &  $P4/mmm$ (123)       & 0.53 &30 (16, 14)   &  $-0.009$& 0.194	& 1.017 & *    \\ %
NS3-04 &  $C2/m$ (12)          & 0.53 &30 (16, 14)   &  $-0.006$& 0.197	& 1.016 & 1.14 \\ %
NS3-05 &  $Cmmm$ (65)          & 0.53 &30 (16, 14)   &  $-0.001$& 0.202	& 1.013 & 0.53 \\ %
NS3-06 &  $P2/c$ (13)          & 0.53 &30 (16, 14)   &  $ 0.000$& 0.204	& 1.014 & 1.20 \\ %
NS3-07 &  $Fmmm$ (69)          & 0.53 &15 (8, 7)     &  $ 0.010$& 0.213	& 1.015 & 2.66 \\ %
NS3-08 &  $Amm2$ (38)          & 0.53 &15 (8, 7)     &  $ 0.026$& 0.229	& 1.192 & 1.09 \\ %
\hline
NS4-01 &  $R\bar{3}m$ (166)    & 0.54 &26 (14, 12)   &  $-0.027$& 0.208	& 1.015 & *    \\ %
NS4-02 &  $C2/m$ (12)          & 0.54 &26 (14, 12)   &  $-0.013$& 0.222	& 1.015 & 0.35 \\ %
NS4-03 &  $I4/mmm$ (139)       & 0.54 &13 (7, 6)     &  $-0.010$& 0.224	& 1.019 & *    \\ %
NS4-04 &  $C2/m$ (12)          & 0.54 &26 (14, 12)   &  $-0.007$& 0.228	& 1.018 & 1.10 \\ %
NS4-05 &  $P2/m$ (10)          & 0.54 &26 (14, 12)   &  $-0.002$& 0.233	& 1.015 & 0.16 \\ %
NS4-06 &  $Immm$ (71)          & 0.54 &13 (7, 6)     &  $ 0.003$& 0.238	& 1.015 & 0.57 \\ %
\hline
NS5-01 &  $R\bar{3}m$ (166)    & 0.55 &22 (12, 10)   &  $-0.035$& 0.243	& 1.019 & *    \\ %
NS5-02 &  $C2/m$ (12)          & 0.55 &33 (18, 15)   &  $-0.027$& 0.251	& 1.023 & *    \\ %
NS5-03 &  $Cm$ (8)             & 0.55 &33 (18, 15)   &  $-0.022$& 0.255	& 1.018 & *    \\ %
NS5-04 &  $Pmm2$ (25)          & 0.55 &33 (18, 15)   &  $-0.022$& 0.255	& 1.020 & *    \\ %
NS5-05 &  $Pmm2$ (25)          & 0.55 &33 (18, 15)   &  $-0.019$& 0.258	& 1.018 & 0.55 \\ %
NS5-06 &  $P4/mmm$ (123)       & 0.55 &11 (6, 5)     &  $-0.012$& 0.265	& 1.022 & *    \\ %
NS5-07 &  $C2/m$ (12)          & 0.55 &11 (6, 5)     &  $ 0.002$& 0.280	& 1.018 & 0.71 \\ %
NS5-08 &  $P2/m$ (10)          & 0.55 &11 (6, 5)     &  $ 0.002$& 0.280	& 1.018 & 0.69 \\ %
NS5-09 &  $C2/m$ (12)          & 0.55 &11 (6, 5)     &  $ 0.003$& 0.280	& 1.018 & 0.92 \\ %
NS5-10 &  $Pmmm$ (47)          & 0.55 &11 (6, 5)     &  $ 0.003$& 0.281	& 1.018 & 0.49 \\ %
\hline
NS6-01 &  $R\bar{3}m$ (166)    & 0.56 &9  (5, 4)     &  $-0.036$& 0.303	& 1.019 & *    \\ %
NS6-02 &  $Imm2$ (44)          & 0.56 &27 (15, 12)   &  $-0.027$& 0.312	& 1.024 & *    \\ %
NS6-03 &  $P2_1/m$ (11)        & 0.56 &18 (10, 8)    &  $-0.017$& 0.322	& 1.021 & 0.68 \\ %
NS6-04 &  $I4/mmm$ (139)       & 0.56 &9  (5, 4)     &  $-0.015$& 0.324	& 1.026 & *    \\ %
NS6-05 &  $C2/m$ (12)          & 0.56 &18 (10, 8)    &  $-0.010$& 0.329	& 1.026 & 0.87 \\ %
NS6-06 &  $Immm$ (71)          & 0.56 &27 (15, 12)   &  $-0.006$& 0.333	& 1.026 & *    \\ %
NS6-07 &  $P2/m$ (10)          & 0.56 &18 (10, 8)    &  $-0.003$& 0.336	& 1.022 & *    \\ %
NS6-08 &  $C2/m$ (12)          & 0.56 &9  (5, 4)     &  $ 0.001$& 0.340	& 1.023 & 0.90 \\ %
NS6-09 &  $Immm$ (71)          & 0.56 &9  (5, 4)     &  $ 0.001$& 0.340	& 1.023 & 0.55 \\ %
\hline
NS7-01 &  $R\bar{3}m$ (166)    & 0.57 &28 (16, 12)   &  $-0.047$& 0.389	& 1.029 & *    \\ %
NS7-02 &  $P2_1/m$ (11)        & 0.57 &28 (16, 12)   &  $-0.044$& 0.392	& 1.029 & *    \\ %
NS7-03 &  $Pmmm$ (47)          & 0.57 &28 (16, 12)   &  $-0.039$& 0.397	& 1.032 & *    \\ %
NS7-04 &  $Cmmm$ (65)          & 0.57 &28 (16, 12)   &  $-0.039$& 0.397	& 1.031 & *    \\ %
NS7-05 &  $C2/m$ (12)          & 0.57 &28 (16, 12)   &  $-0.037$& 0.399	& 1.035 & *    \\ %
NS7-06 &  $Cccm$ (66)          & 0.57 &28 (16, 12)   &  $-0.031$& 0.405	& 1.029 & *    \\ %
NS7-07 &  $P2_1/m$ (11)        & 0.57 &14 (8, 6)     &  $-0.025$& 0.411	& 1.027 & 0.26 \\ %
NS7-08 &  $P2_1/m$ (11)        & 0.57 &14 (8, 6)     &  $-0.023$& 0.413	& 1.028 & 0.49 \\ %
NS7-09 &  $P4/mmm$ (123)       & 0.57 &7  (4, 3)     &  $-0.020$& 0.416	& 1.033 & *    \\ %
\arrayrulecolor{black}\hline
\end{tabular}
\end{ruledtabular}
\end{table*} 

The dynamical stability of these structures has also been verified 
by phonon calculations at the DFT level. A variety of supercells 
were used for various unit cell sizes to verify the convergence 
in the phonon DOSs and force constants. 
The phonon calculations show that all the non-stoichiometric 
structures reported in this paper are dynamically stable.
Note that we neglect the Born effective charge in the phonon calculations
thus the LO-TO splitting is lacking in the phonon dispersion curves.
The electronic band structures and band gaps for the non-stoichiometric 
phases were calculated at the PBE level as indicated in \cref{tab:nonst}. 
In contrast to the stoichiometric polymorphs which have large gaps, 
the energy band gap of suboxide non-stoichiometric phases of MgO is 
small and $23$ structures have even a pseudogap. In fact, removing
O$^{2-}$ anions results in new states in the band gap of the structures
due to the extra electrons that trapped in the oxygen vacancies or
the so-called F-centers~\cite{Pacchioni_2008}.
This indicates that the most of the suboxide non-stoichiometric 
polymorphs of MgO have physical properties like semimetals or 
perhaps self-doped semiconductors.

\begin{figure*}[ht]
    \centering
        \includegraphics[width=2.00\columnwidth]{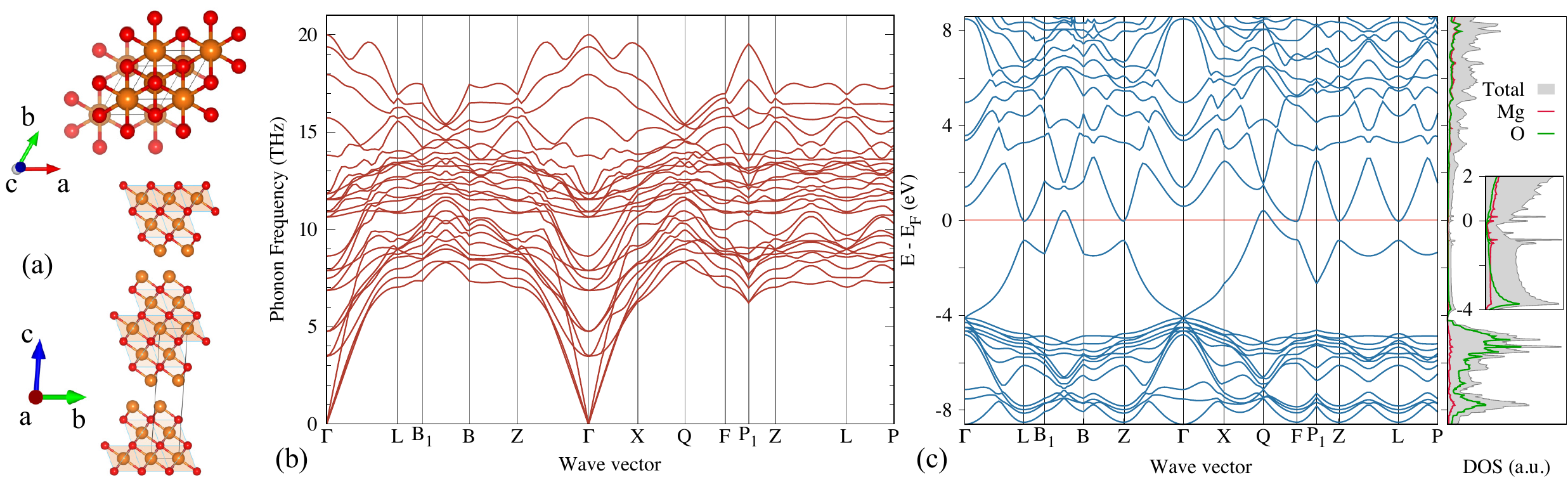}
    \caption{(a) Crystal structure of NS6-01
    viewed along the $[111]$ direction  (top) and
    the $[100]$ direction (bottom) of the RS phase . (b) PBE phonon dispersion
    (by neglecting LO-TO splitting)
    and (c) PBE electronic band structure together with 
    the total and projected density of states of the polymorph.
    The inset magnifies the density of states near the Fermi energy.}
    \label{fig:str_spg166}
\end{figure*}

As our results in \cref{tab:nonst} indicate, 
the GM structures of suboxide non-stoichiometric compositions
i.e. NS1-01, NS2-01, NS3-01, NS4-01, NS5-01, NS6-01, and NS7-01, 
are similar with space groups $R\bar{3}m$ or $P\bar{3}m1$.
Indeed, they are formed by ordering of the oxygen vacancies on 
the $(111)$ plane of the RS phase.
These structures which are indicated by the red color in \cref{fig:energy_x}, 
have the same structural motifs as depicted in 
\cref{fig:polyhedra}(f) for NS7-01 and \cref{fig:str_spg166}(a) for NS6-01.

In the $xy$-plane the GM structures coincide with the RS phase and along 
the $z$-direction they are composed of rock salt $(111)$ slabs.
The top and bottom layers in these slabs are Mg atoms due 
to the oxygen-vacancy ordering in the $(111)$ plane. 
These slabs are stacked on top of each other without any bonding between them, 
and only the number of atoms per primitive cell has changed for each 
composition $x$. For instance, NS6-01 has 5 magnesium atoms and 4 oxygen atoms
which determine the thickness of the slabs in this structure.  
This is illustrated in \cref{fig:polyhedra}(f) and \cref{fig:str_spg166}(a).
Although the GM structure of two compositions $x=0.51$ and $0.53$ have 
symmetries which are different from the others, these structures are 
very similar since the belong to the same point group $\bar{3}2/m$.

\begin{figure}[ht]
    \centering
    \includegraphics[width=1.00\columnwidth]{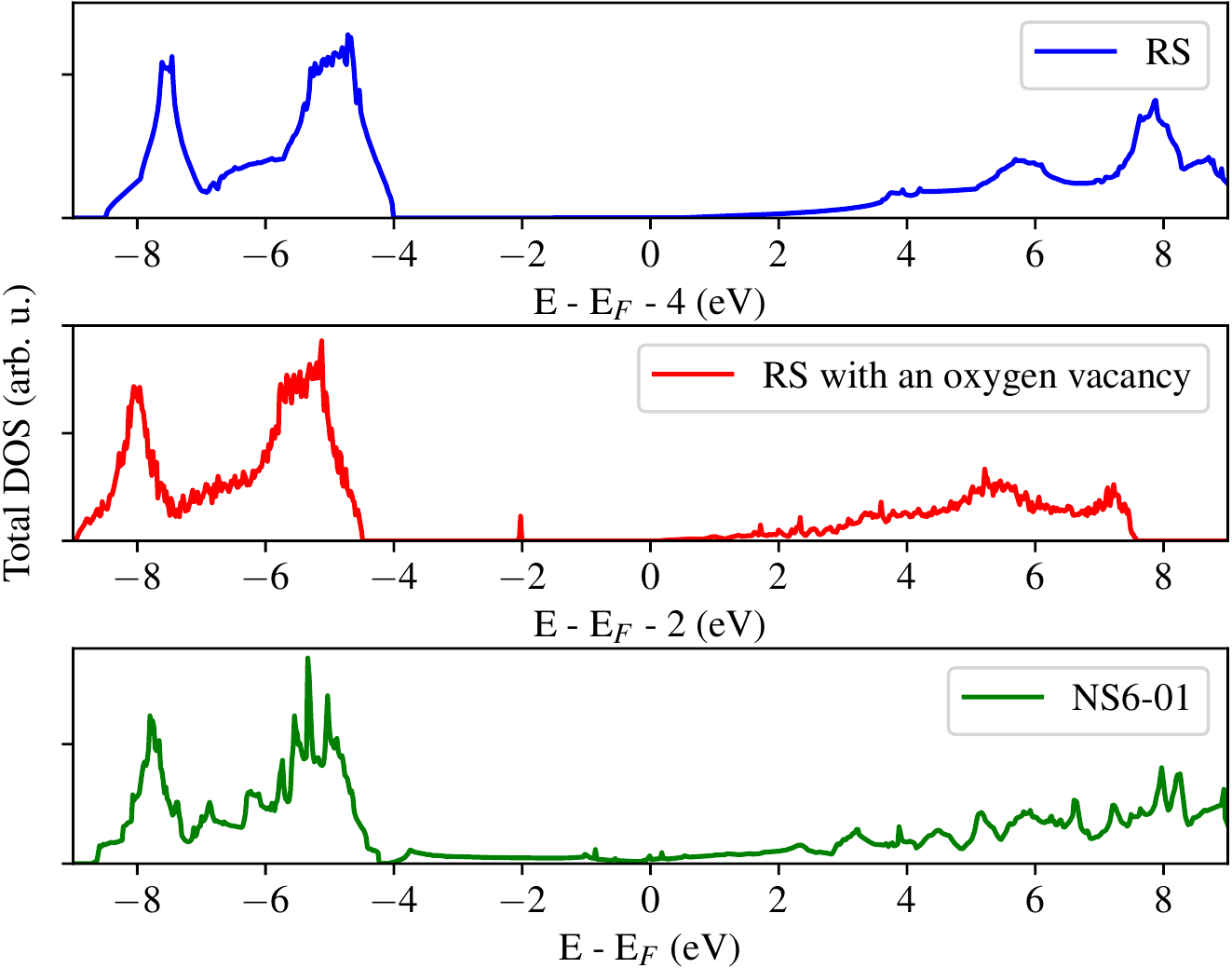}
    \caption{Comparison of the electronic densities of states of the RS phase 
    with and without an oxygen and of phase NS6-01.
    For better visibility, the Fermi energy of the first two structures
    are shifted by $-4$ and $-2$ eV, respectively.}
    \label{fig:dos_spg166}
\end{figure}

Our calculations demonstrate that these GM structures have a pseudogap
in contrast to the large band gap of the perfect RS structure.
The density of states also differs from the one of a perfect RS structure 
with a single vacancy. In the latter case there is a single defect level peak in the 
density of states whereas in the former case there is a low but uniform 
density of states in the original gap (see \cref{fig:dos_spg166}).
This small and quite uniform density of states arises from the 
strong dispersion of the electronic band in the original RS band gap is shown in
\cref{fig:str_spg166}(c).
For this polymorph, we also performed the electronic structure calculations
with the hybrid functional PBE0~\cite{Perdew_1996,Adamo_1999}. 
Although hybrid functionals usually give a better description of band 
gap, close to the experimental values, it yields the same results as PBE for this structure. 
Therefore, in the GM structure of each composition the Fermi level is crossed by
the electronic states which can give rise to interesting electronic properties.

The phonon dispersions of NS6-01 are illustrated in \cref{fig:str_spg166}(b)
which shows dynamical stability of these structures. 
Note that for the phonon and electronic band structure calculations 
of these GM structures, we have also applied Van der Waals corrections.

We also investigated the lattice thermal properties since the CENT potential
can give a good and reliable description of the third-order IFCs.
Although in high purity metals, the electronic part of the thermal conductivity
dominates~\cite{Yao_2017}, we expect that due to the small amount of electrons which 
lie above the pseudo gap in the GM structures, the lattice thermal conductivity
should be important for these structures.

\begin{figure}[ht]
    \centering
    \includegraphics[width=1.00\columnwidth]{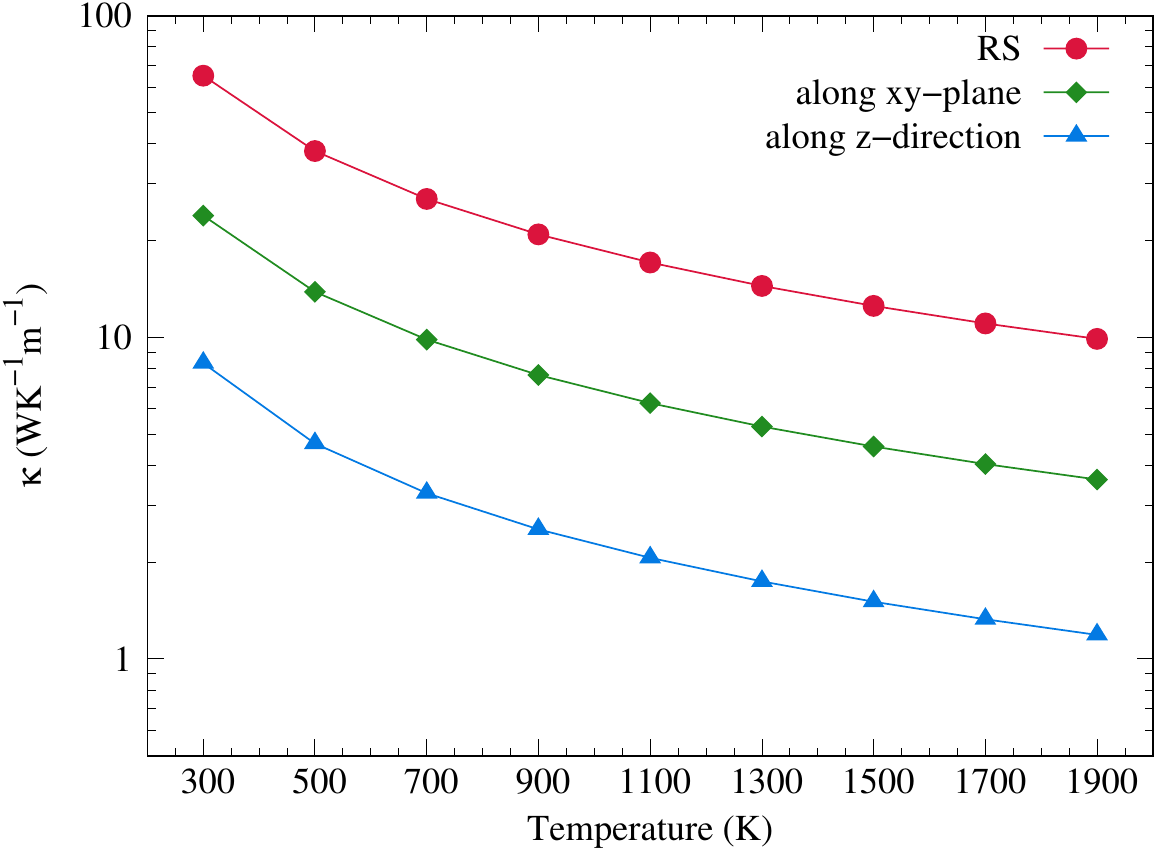}
    \caption{Comparison of thermal conductivity of the non-stoichiometric structure 
    NS6-01 and the RS phase.}
    \label{fig:TC_spg166}
\end{figure}

For large unit cells, assessing the lattice 
thermal conductivity from first principles is computationally prohibitive. 
We therefore resort to employ only the CENT potential
to compute the anharmonic interatomic force constants.
We calculate the thermal conductivity of NS6-01 using a 
supercell of $4\times4\times2$ including $288$ atoms. 
We only consider atomic interactions up to the sixteenth nearest 
neighbors for which the lattice thermal conductivity is converged.
To compute the anharmonic IFCs for this system with $9$ atoms 
per primitive cell, the atomic forces on $4576$ displaced structure 
are required. 
In addition to the supercell size and the converged cutoff value, 
we tested the other parameters, e.g.
the number of k-points along each axis in the reciprocal space and
the scale parameter for Gaussian smearing, that are necessary
to obtain a converged thermal conductivity with the approach implemented 
in the ShengBTE code.
The results are depicted in \cref{fig:TC_spg166} in which 
the thermal conductivities of NS6-01 along the $x-y$ plane and 
$z$-direction are compared with the RS structure.
The thermal conductivity in the $z$-direction decreases 
significantly up to one order of magnitude in comparison with the RS phase. 
According to Slack's rule, the structures with a large number of atoms per 
primitive cell have very low values of thermal conductivity
since optical modes play a key role in phonon scattering
~\cite{Slack_1973,Eucken1928May}. 
On the other hand, the slabs geometry could be the other reason 
why only little heat is carried along the $z$-direction.
Therefore, this amount of reduction in the thermal conductivity 
would not be surprising.
However, such a drastic decrease in the thermal conductivity 
suggests that the non-stoichiometric phases of MgO could be potentially 
good candidates to use in several applications similar as alternatives to 
chemical compounds of group IV-VI elements. 
Also, these deviations of the electronic and thermal properties from 
the stoichiometric structures confirm that the GM structures 
of the non-stoichiometric suboxides are similar to the Magn\'eli-type
oxides which have been reported for some materials~\cite{Harada_2010,Kieslich_2014}.

\section{Conclusion}\label{sec:conclusion}
We constructed a highly transferable CENT potential to explore the energy 
landscape of MgO compounds for both stoichiometric and 
non-stoichiometric compositions. This neural network potential is validated 
by various challenging tests such as predicting the PES of clusters and 
crystals of MgO as well as computing phonons and thermal transport properties.
During our search, we found all the neutral (MgO)$_n$ ($n=2-40$) 
clusters and stoichiometric bulk structures of MgO 
which have been reported in the literature. Also, the phonon dispersion and 
thermal conductivity obtained for RS structures is in good agreement
with the DFT results and previous theoretical and experimental works.

Subsequently, we systematically explored the phase diagram of 
Mg$_x$O$_{1-x}$ with a focus on the near-stoichiometric range. 
We found new low energy stoichiometric 
polymorphs which are modifications of the rock salt phase and studied $48$ 
new metastable structures with non-stoichiometric compositions. 
We demonstrated that there is a dense spectrum of 
polymorphs for each composition of $x$ ($0.51\leq x\leq0.57$). 
They all lie in a narrow energy range and are energetically more 
favorable than RS with one to four oxygen defects. This energy lowering is 
due to an oxygen-vacancy ordering along planes and lines.

Also, our results show that the GM structure of each composition 
is a structure in which the oxygen vacancies are located in $(111)$ planes. 
Therefore, these structures are composed of stacked slabs of the RS structure 
in the direction $[111]$ which terminate with Mg atoms.
The number of layers of each slab then depends on the number of atoms in 
their unit cell which it increases with decreasing $x$. 
Although the GM structures have the same structural motifs,
their symmetry groups are slightly different i.e. $P\bar{3}m1$ 
for two compounds ($x=0.51$ and $0.53$) and $R\bar{3}m$ for the others. 
They have peculiar thermal and electronic properties like Magn\'eli-type 
oxides. Thus it would be of interest to synthesize them for practical applications.

\section*{Acknowledgments}
The authors gratefully acknowledge support from the scientific computing core (sciCORE)
facility at the University of Basel.
Also, S. A. G. and H. T. thank Maximilian Amsler for valuable expert discussions 
and proofreading the manuscript.

\bibliography{main}

\end{document}